# Dissipative phases across the superconductor-to-insulator transition

F. Couëdo, O. Crauste, A.A. Drillien, V. Humbert, L. Bergé,
C.A. Marrache-Kikuchi* and L. Dumoulin

CSNSM, Univ. Paris-Sud, CNRS/IN2P3, Université Paris-Saclay, 91405 Orsay, France

*Corresponding author (email: claire.marrache@csnsm.in2p3.fr)

**Competing phenomena in low dimensional systems can generate exotic electronic phases, either through symmetry breaking or a non-trivial topology. In two-dimensional (2D) systems, the interplay between superfluidity, disorder and repulsive interactions is especially fruitful in this respect although both the exact nature of the phases and the microscopic processes at play are still open questions. In particular, in 2D, once superconductivity is destroyed by disorder, an insulating ground state is expected to emerge, as a result of a direct superconductor-to-insulator quantum phase transition. In such systems, no metallic state is theoretically expected to survive to the slightest disorder. Here we map out the phase diagram of amorphous NbSi thin films as functions of disorder and film thickness, with two metallic phases in between the superconducting and insulating ones. These two dissipative states, defined by a resistance which extrapolates to a finite value in the zero temperature limit, each bear a specific dependence on disorder. We argue that they originate from an inhomogeneous destruction of superconductivity, even if the system is morphologically homogeneous. Our results suggest that superconducting fluctuations can favor metallic states that would not otherwise exist.**

In 2D systems, disorder induces quantum interferences between electronic wave functions, eventually leading to their localization. No matter how weak the disorder, Anderson localization prevents the diffusion of the electronic motion so that no 2D metal can exist [1]. However, in the presence of competing orders, new electronic ground states can prevail [2]. In the case of thin disordered superconducting films, when Cooper pairing, competing with Coulomb repulsion, is not sufficient to establish a long-range coherence between localized states, an insulator is expected to arise from the superconducting ground state. This Superconductor-to-Insulator Transition (SIT) is either explained by amplitude or phase fluctuations of the superconducting order parameter, and conventional theories do not allow for any intervening metallic state [3]. Experimentally however, dissipative behaviors have been observed [4-11]. Several hypotheses have been put forward to explain these, amongst which: vortices-induced dissipation [4,5]; a coupling to a dissipative bath [6]; the existence of a quasi-2D metal [12]; the competition between Josephson coupling and the charging energy [13]; important Coulomb interactions [14]. Moreover, in the vicinity of the SIT, the focus has also recently been set on fluctuations as a possible cause for inhomogeneous electronic phases [15-16]. The nature and origin of such metallic-like behaviors is thus still debated. Whether they live on at *T* = 0 and constitute one of the system's ground states is also an important issue that needs to be solved. In order to gain more insight on these questions, a systematic quantitative analysis of these dissipative phases would be extremely profitable.

Thin metal alloy films constitute particularly well-suited systems for this study. Indeed, in compounds such as a-$Nb_xSi_{1-x}$ (a-NbSi) which we consider, disorder can be progressively increased either by a reduction of the sample thickness or by a variation in stoichiometry, but also through a thermal treatment [17-18]. In this letter, we present the phase diagram of a-NbSi thin films across the disorder-induced SIT, along with the characterization of corresponding phases. Once superconductivity is destroyed, we surprisingly identify two metallic regimes – that we have called "Metal 1" and "Metal 2". The insulating phase subsequently appears for films of normal conductivity smaller than $e^2/h$. This suggests that, in this system, the SIT and the Metal-to-Insulator Transition (MIT) happen successively. The Metal 1 can be interpreted as emerging from superconducting fluctuations, whereas the Metal 2 should be of fermionic nature.

In this study, we have considered 40 samples with compositions $x$ from 8.5 to 18.5 %, thicknesses $d$ from 4 to 50 nm, and submitted to heat treatments at temperatures $\theta_{\text{ht}}$ from 70 to 250 °C. The effects of these disorder-changing parameters ($x, d, \theta_{\text{ht}}$) on the properties of superconducting a-NbSi thin films have been described elsewhere [17]. We will here examine the transport features of the films as a function of $\sigma_{\text{N}}$, the normal state conductivity at $T = 500$ mK, taken as a measure of disorder. All resistances (conductances) will be plotted in units of the quantum of resistance (conductance) $h/e^2$ ($e^2/h$). The films are continuous, uniformly disordered and amorphous up to $\theta_{\text{ht}} = 500$ °C. The deposition and heat treatment procedures are described in the Methods section.

When superconducting, the samples are either 2D or in the quasi-2D limit, i.e. $d \lesssim \xi$, where $\xi$ is the superconducting coherence length. The films were measured using standard low frequency transport measurements at very low temperature (see Methods).

All films fall into one of four categories, identified by their low temperature transport characteristics. We will identify each one on a specific example – a 23-nm-thick a-Nb$_{13.5}$Si$_{86.5}$ film that has been thermally treated –. We will then quantitatively analyze each ground state and show how their features evolve with disorder.

The low temperature transport properties of the 23-nm-thick a-Nb$_{13.5}$Si$_{86.5}$ film are represented in Fig. 1a. The as-deposited film is superconducting, with a well-defined zero resistance state below the critical temperature $T_c = 50$ mK. As is usual, the superconducting phase presents a positive Temperature Coefficient of Resistance ($TCR = \frac{dR}{dT}$) at low temperature. After thermal treatment, $\sigma_N$ decreases, signaling an enhanced effective disorder. Up to $\theta_{ht} = 140$ °C, the low temperature $TCR$ stays positive, but the sheet resistance $R_{min}$ measured at $T = 10$ mK is finite. For $\theta_{ht} = 110$°C, the resistance even saturates at low temperature. The sample then is in a phase that we shall call Metal 1. For $\theta_{ht} > 140$ °C, the sample is characterized by a negative $TCR$ and a finite $R_{min}$, corresponding to a phase we have named Metal 2. All finite resistances measured at our lowest temperature have been checked to be intrinsic and not due to experimental artefacts (see Methods). At even larger disorder, the system eventually becomes insulating. This phase can be described by a negative *TCR* and a diverging low temperature resistance (not reached in the data set shown in Fig. 1a).

Let us analyze each phase more quantitatively. First, the superconducting state is characterized by two temperature scales. Starting from the high temperature regime, the resistance abruptly drops at $T_{c0}$, defined by the temperature at which the derivative is maximum ($\left(\frac{dR}{dT}\right)_{T=T_{c0}} = max\left(\frac{dR}{dT}\right)$). For low disordered films, this energy scale corresponds to the characteristic superconducting energy (see Methods and supplementary materials). The resistance then becomes null below the superconducting critical temperature $T_c$, indicating the establishment of global phase coherence. The evolution of these two characteristic temperatures with disorder is shown in Fig. 1b for 23-nm-thick samples. At low disorder, the two energy scales evolve in a similar manner. However, at a first critical conductivity $\sigma_{c1}$, $T_c$ drops to zero, signaling the destruction of the macroscopic phase coherence, while $T_{c0}$ continues to be finite. $\sigma_{c1}$ therefore marks the transition between the superconducting and the Metal 1 phases. As will be seen in the phase diagram (Fig. 4), $\sigma_{c1}$ depends linearly on the thickness $d$ and, in the strictly 2D limit, extrapolates to $\sigma_Q = 4e^2/h$. In the Metal 1 phase, the $R(T)$ characteristics abruptly drop below $T_{c0}$, but, while always maintaining a positive $TCR$, extrapolate to a finite value at $T = 0$. We have estimated this residual resistance to be close to $R_{\min}$, the resistance measured at $T = 10$ mK. The evolution of $T_{c0}$, $R_{\min}$, and of the sign of the $TCR$ – all three characteristic of the Metal 1 phase – with disorder is given in Fig. 1b for the 23-nm-thick a-Nb$_{13.5}$Si$_{86.5}$ film. Although, in this phase, $T_{c0}$ no longer represents the superconducting critical temperature, its continuous evolution from the superconductor into the Metal 1 regime suggests the importance of superconducting fluctuations in this latter state. As can be seen, $R_{\min}$ rapidly increases with increasing disorder, until, at a second critical conductivity $\sigma_{c2}$, the three quantities reflect a change of regime : the evolution of the residual resistance slows down, $T_{c0}$ extrapolates to zero and the $TCR$ changes sign. For *all* measured samples, the modifications in the behavior of the $TCR$, of $R_{\min}$ and of $T_{c0}$ simultaneously occur at $\sigma_N = 1/R_{\min}$, i.e. for a sample which would have a constant conductivity from $T = 10$ mK to $T$ = 500 mK. However $\sigma_{c2}$ is not universal and varies with the film thickness as can be seen in Fig. 4.

Let us now specifically consider the Metal 2 phase. Figure 2a shows the evolution of the transport characteristics of a 5-nm-thick a-Nb$_{13.5}$Si$_{86.5}$ film with disorder. All curves show the same qualitative features: from the high temperature regime, the resistance first progressively increases as is expected on the insulating side of the SIT. However, below a characteristic temperature $T_\text{sat}$, the concavity of the $R(T)$ curve changes, while maintaining a negative $TCR$, and levels off at low temperature at $R_\text{min}$. $T_\text{sat}$ can then be defined as the highest temperature at which $R_\text{min}$ is attained within 1%. This temperature is characteristic of the energy scale at which the saturation of the resistance sets in. The evolution of $T_\text{sat}$ with disorder is plotted in Fig. 2b for the 5-nm-thick a-Nb$_{13.5}$Si$_{86..5}$ film: as disorder is increased, the temperature range over which $R \simeq R_\text{min}$ shrinks, while the product $R_\text{min}\sigma_\text{N}$ becomes larger, showing a progressive weakening of the metallic state. In Fig. 2c is plotted the reduction of the residual conductivity $\sigma_\text{min} = 1/R_\text{min}$ as the disorder level is increased, for *all* samples in the Metal 2 regime. The evolution is universal, thus showing that the sole knowledge of normal state conductivity $\sigma_\text{N}$ fully determines the value of $\sigma_\text{min}$, whatever the experimental disorder-tuning parameter ($x$, $\theta_\text{ht}$ or $d$). This second metallic regime vanishes close to $\sigma_\text{N} \simeq e^2/h$.

For films such that $\sigma_\text{N} < e^2/h$, transport properties are consistent with an insulator typical of the SIT: as can be seen in Fig. 3a, they are well described by a phenomenological Arrhenius law: $R(T) = R_0 \exp(T_0/T)$. The evolution with disorder of the characteristic temperature $T_0$ and the corresponding $R_0$ is displayed in Fig. 3b. Like $\sigma_\text{min}$ in the previous Metal 2 phase, $T_0$ decreases universally as $\sigma_\text{N}$ increases, whatever the disorder-tuning parameter. Both characteristic parameters concurrently vanish at a third critical conductivity $\sigma_{c3} = (1.8 \pm 0.8)\, e^2/h$.

Although the microscopic nature of these dissipative regimes is unclear for now, one can formulate some hypotheses. The continuous evolution of $T_{c0}$ from the superconducting phase into the Metal 1, where its value remains finite, assesses the importance of superconducting fluctuations in this first metallic regime. This argument is also supported by the thickness-dependence of all relevant quantities ($T_{c0}$, $R_\text{min}$, $\sigma_{c1}$, and $\sigma_{c2}$) in this phase. Indeed, a previous work had evidenced the specific role of $d$, as opposed to $\theta_\text{ht}$ or $x$, on the superconducting properties of a-NbSi films [17]. This distinct role is also found in the Metal 1 phase, but not in the Metal 2 or insulating regimes. The Metal 1 could then correspond to a phase where short living Cooper pairs survive locally to the destruction of global superconductivity. In the literature, one can find both theoretical predictions and experimental evidence of such an inhomogeneous electronic phase with non-zero pairing amplitude, coming from the destruction of superconductivity via phase fluctuations [8-9,13,15,19]. However these works considered systems with built-in inhomogeneities and cannot, rigorously, be applied to our system. Moreover, and although numerous experimental and theoretical works dealt with the possibility of a metallic state as a result of the interaction between disorder and superconductivity [4-13, 15-16], to the best of our knowledge, there has been no prediction of two distinct dissipative regimes in morphologically homogeneous systems at zero magnetic field. In the Metal 2 phase, by contrast, the transport characteristics seem to be only governed by the normal conductivity $\sigma_N$. This signals a transition or a crossover to a quasiparticules-dominated transport regime. In this phase, the saturation of the resistance at low temperature could be explained by a parallel channel of conduction short-circuiting the localized fermions [20]. The origin or nature of this additional channel, whether of quantum metal or incoherent Cooper pairs, is still to be determined, but also supports the inhomogeneous nature of electronic transport in this system.

All our results can be summarized in the phase diagram shown in Fig. 4, where the different regimes are represented as a function of disorder and film thickness. As can be seen, at lower thicknesses, the superconducting state occupies a larger portion of the phase diagram. Although counter-intuitive, this reflects how enhanced superconducting fluctuations may actually help the establishment of global phase coherence. In the $d \to 0$ limit, Metal 2 vanishes, as could be expected in the case of a fermionic metal [1] where the least amount of disorder is sufficient to localize the electrons at low enough temperature. The insulating phase sets in at the critical conductivity $\sigma_{c3} \simeq e^2/h$, independently of the film thickness. This level of disorder corresponds to the Ioffe-Regel criterion $k_F l \simeq 1$ at which the MIT is expected in 3D [21-22]. This criterion has also been observed in 2D electron gases at the interface of semiconductor heterostructures [23]. The most surprising feature of this phase diagram is the appearance of the Metal 1 phase, which vanishes in the 3D limit but survives in the strictly two-dimensional limit, insofar as our results can be extrapolated. It grows from superconductivity and accompanies its evolution down to the lowest thicknesses. This intriguing dissipative state shows how quantum fluctuations do not only destroy ordering but may also significantly contribute to the appearance of new electronic phases.

**Methods**

**Sample deposition and heat treatment procedure**

$a$-NbSi films have been prepared at room temperature and under ultrahigh vacuum (typically a few $10^{-8}$ mbar) by electron beam co-deposition of Nb and Si, at a rate of the order of $1\,\text{Å}.\text{s}^{-1}$. The evaporation rates of each source were monitored in situ by a dedicated set of piezoelectric quartz crystals in order to precisely monitor the composition and the thickness of the films during the deposition. These were also corroborated ex situ by Rutherford backscattering spectroscopy (RBS) measurements.

The films were deposited onto sapphire substrates coated with a 25-nm-thick SiO underlayer designed to smooth the substrate surface. The samples were subsequently protected from oxidation by a 25-nm-thick SiO overlayer. Similar films have been measured to be continuous, amorphous, and structurally non-granular at least down to a thickness $d = 25$ nm [17].

The as-deposited films are considered to have experienced a thermal treatment at $\theta_{\text{ht}} = 70\,°C$, due to heating during the deposition process. This was confirmed by low-temperature measurements of the conductivity [17]. Subsequent heat treatments have been performed for 1 h, under a flowing $N_2$ atmosphere. To prevent any thermal shock, the samples were then slowly cooled back down to room temperature. TEM measurements were performed on similar films that were annealed in situ from $\theta_{\text{ht}} = 70\,°C$ to 700 °C: no structural nor composition changes have been observed up to an annealing temperature of $\theta_{\text{ht}} = 500\,°C$ [17].

**Estimation of the superconducting coherence length**

A *minimal* estimate of the superconducting coherence length can be derived from the critical temperature $T_{c,\text{bulk}}$ and normal conductivity $\sigma_{N,\text{bulk}}$ of a bulk sample of same stoichiometry, using Gor'kov's development of the Ginzburg-Landau theory in the dirty limit : $\xi = 0.36\sqrt{\frac{3}{2}\frac{3\hbar D}{k_B T_{c,\text{bulk}}}}$ with D = 0.6 cm².s⁻¹, extracted from critical field measurements [24]. For $x = 13.5\%$ ($T_{c,\text{bulk}} = 146$ mK), we have $\xi = 43$ nm and for $x = 18\%$ ($T_{c,\text{bulk}}$ = 965 mK), $\xi = 17$ nm [17]. All the considered samples have $d < \xi$ and can therefore be considered to be 2D from the point of view of superconductivity.

**Transport measurements**

Transport measurements were carried out down to 10 mK in a dilution refrigerator, using a resistance measurement bridge and standard ac lock-in detection techniques. The applied bias has been checked to be sufficiently low to be in the ohmic regime, or, when superconducting, below the critical current. All electrical leads are thermalized at the mixing chamber and at the different stages of the cryostat. They are also filtered from RF at room temperature.

To ensure the intrinsic nature of the low temperature dissipative regimes, we have checked that the measured residual resistances $R_{\min}$ do not depend on the sample area, nor on its volume or geometry. Note that these dissipative states have also been observed in the case of a single sample that has been progressively annealed, and measured in the same measurement environment.

**Determination of $T_{c0}$**

The characteristic temperature $T_{c0}$ is defined by the temperature at which the derivative is maximum ($\left(\frac{dR}{dT}\right)_{T=T_{c0}} = max\left(\frac{dR}{dT}\right)$). This criterion agrees well with the temperature extracted from superconducting fluctuations analysis close to the superconducting transition. The comparison of the two methods is presented in the supplementary material.

**Estimation of $R_{min}$**

We have taken $R_{min} = R(T = 10 \text{ mK})$ as an estimate of the residual resistance at zero temperature. For samples that present resistance saturations at low temperature (both in the "Metal 1" and "Metal 2" regimes), the extrapolation of the resistance at zero temperature is equal (within 1%) to $R_{min}$. For samples that have not reached the saturating temperature (sample with $\theta_{ht}$= 130 °C in figure 1a for example), equating the zero-temperature residual resistance to $R_{min}$ is an approximation. The corresponding error can be estimated by taking the low temperature slope of the $R(T)$ curve and extrapolate it to $T = 0$. The error has been found to be smaller than 2% (except for $\theta_{ht}$ = 130°C: 18%).

**Estimation of $\sigma_{c3}$**

$\sigma_{c3}$ can estimated by two different methods:

- Either by fitting the $T_0(\sigma_N)$ curve of fig 3.b. This gives $\sigma_{c3} \approx 1.3$.
- Or by taking the median point between the last metallic sample ($\sigma_N = 2.6 \frac{e^2}{h}$) and the first insulating one ($\sigma_N = 0.98 \frac{e^2}{h}$) which gives $\sigma_{c3} = 1.8 \pm 0.8$.

It is the latter criterion we have considered to draw the phase diagram in Fig. 4 since the corresponding error bars encompass the first two criteria.

**References**

[1] Abrahams, E., Anderson, P.W., Licciardello, D. C. & Ramakrishnan, T. V. Scaling theory of localization: absence of quantum diffusion in two dimensions. *Phys. Rev. Lett.* **42,** 673 (1979).

[2] Sachdev, S. *Quantum phase transitions* (Cambridge Univ. Press, 1999).

[3] Dobrosavljevic, V., Trivedi, N. & Valles Jr., J.M. *Conductor Insulator Quantum Phase Transitions* (Oxford Univ. Press, 2012).

[4] Lin, Y.-H., Nelson, J. & Goldman, A. M. Suppression of the Berezinskii-Kosterlitz-Thouless transition in 2D superconductors by macroscopic quantum tunneling. *Phys. Rev. Lett.* **109,** 017002 (2012).

**Acknowledgements**

This work has been partially supported by the ANR (grant No. ANR-06-BLAN-0326) and by the Triangle de la Physique (grant No. 2009-019T-TSI2D). We thank L. Benfatto, K. S. Tikhonov and M.V. Feigel'man for stimulating discussions.

**Author contributions**

L.B. prepared the a-NbSi samples. F.C., O.C., A.D., V.H. and C.M.K. carried out the experiments. F.C. and C.M.K. analyzed the data. F.C. and C.M.K. wrote the manuscript and V.H., O.C. and L.D. commented on it. C.M.K. and L.D. initiated this work.

**Competing financial interests**

The authors declare no competing financial interests.

**Figure 1| Superconductor- Metal Transitions**. **a,** Sheet resistance as a function of temperature for a 23-nm-thick a-$Nb_{13.5}Si_{86.5}$ film at annealing temperatures $\theta_{ht}$ ranging from 70 (as-deposited) to 250°C. Left inset: blow-up around the $\theta_{ht} = 150°C$ curve. The horizontal dashed line is a guide for the eyes. Right inset: low temperature saturation regime for $\theta_{ht}$ = 110°C. The horizontal dashed line indicates $R_{min}$. **b,** Evolution of $T_c$ and $T_{c0}$ as a function of the normal state conductivity $\sigma_N$, in the superconducting and Metal 1 regimes, for 23-nm-thick samples (red and black dots respectively). Evolution of the residual resistance $R_{min}$ as a function of the normal state conductivity $\sigma_N$ (blue squares), the blue dashed line corresponds to $R_{min} = 1/\sigma_N$. This illustrates the fact that the transition between the Metal 1 and the Metal 2 phases occurs precisely when $1/\sigma_N = R_{min}$. The vertical pink line corresponds to the change of sign of the $TCR = \frac{dR}{dT}$. Errors bars correspond to the last and first measured samples with negative and positive *TCR.*

**Figure 2|The Metal 2 phase. a,** Sheet resistance as a function of temperature for a 5-nm-thick a-$Nb_{13.5}Si_{86.5}$ film where the disorder is tuned through thermal treatment. Each color corresponds to a different annealing temperature, varied from 70 °C (as deposited film) to 230 °C, by step of 20 °C. **b,** Evolution of the characteristic temperature $T_{sat}$ for the same film, as a function of the normal conductivity $\sigma_N$. **c,** Residual conductivity $\sigma_{min} = 1/R_{min}$ as a function of the normal conductivity $\sigma_N$ for a-NbSi films, with niobium composition $x$ ranging from 10 to 18.6 %, thicknesses $d$ ranging from 3 to 25 nm and heat treatment temperatures up to $\theta_{ht} = 250$ °C. Each symbol represents one film thickness and each color one composition at different annealing temperatures. The dashed line is a guide to the eyes. The dashed line represents the asymptote $\sigma_{min} = \sigma_N$, reached in the limit $\sigma_N \gg e^2/h$.

**Figure 3| The insulating phase. a,** Sheet resistance as a function of the inverse of the temperature for all insulating samples ($8.5\% < x < 10\%$ and $12.5 < d < 26.5$ nm). For each sample, the normal conductivity $\sigma_N$ is given in units of $e^2/h$. Dotted lines are fits to the Arrhenius law. **b,** Evolution of Arrhenius characteristic temperature $T_0$ (red squares) and the residual conductivity $\sigma_{min} = 1/R_{min}$

(black dots) as a function of the normal conductivity $\sigma_N$. The vertical orange line shows the critical disorder $\sigma_{c3}$. Inset: Evolution of Arrhenius characteristic resistance $R_0$ as a function of the normal conductivity $\sigma_N$. The full line is a guide for the eyes.

**Figure 4| Phase diagram of a-NbSi thin films.** Evolution of the film thickness $d$ as a function of the normal conductivity $\sigma_N$. The color points correspond to the critical conductivities separating the different regimes described in the text. Lines are guides for the eyes. The lowest error bars have been determined using the two samples closest to the transition. As a function of $\sigma_N$, the system evolves from the superconducting (SC) phase to the Metal 1 (M1), the Metal 2 (M2) and finally reaches the insulating phase (I). The critical conductivity for the Superconductor - Metal 1 transition extrapolates to $\sigma_Q = 4e^2/h$ in the limit $d \to 0$.

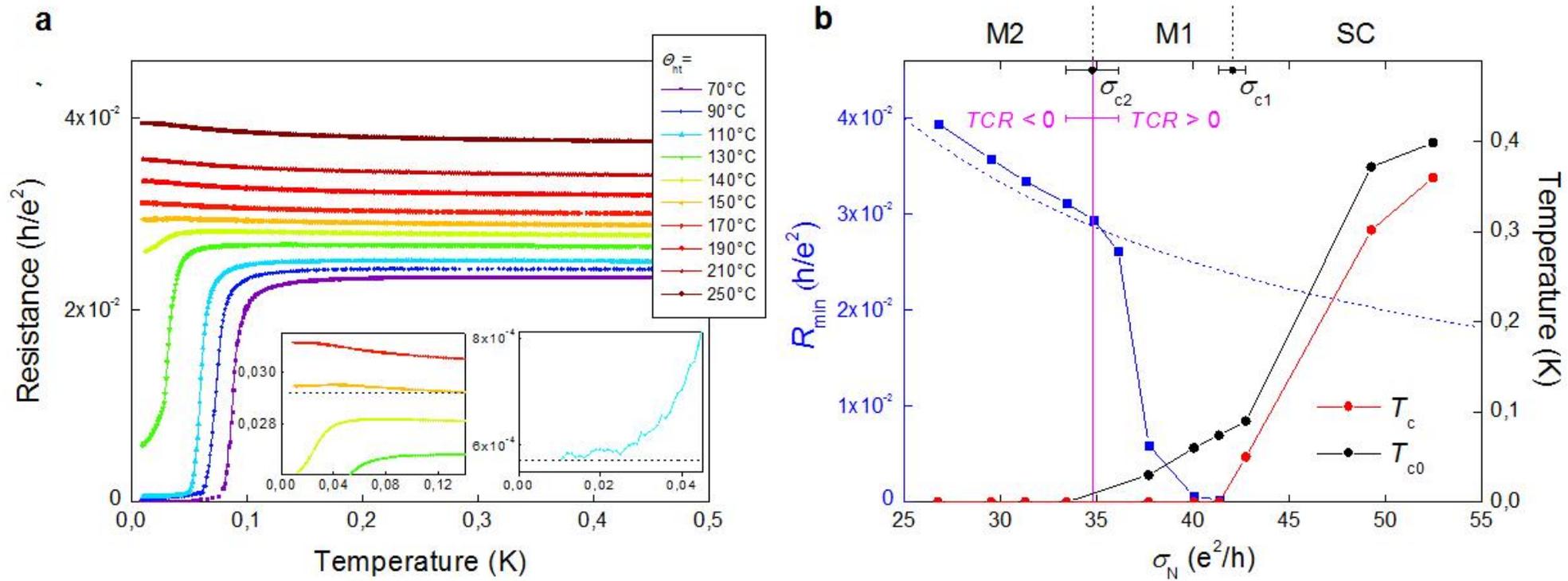

**Fig. 1**

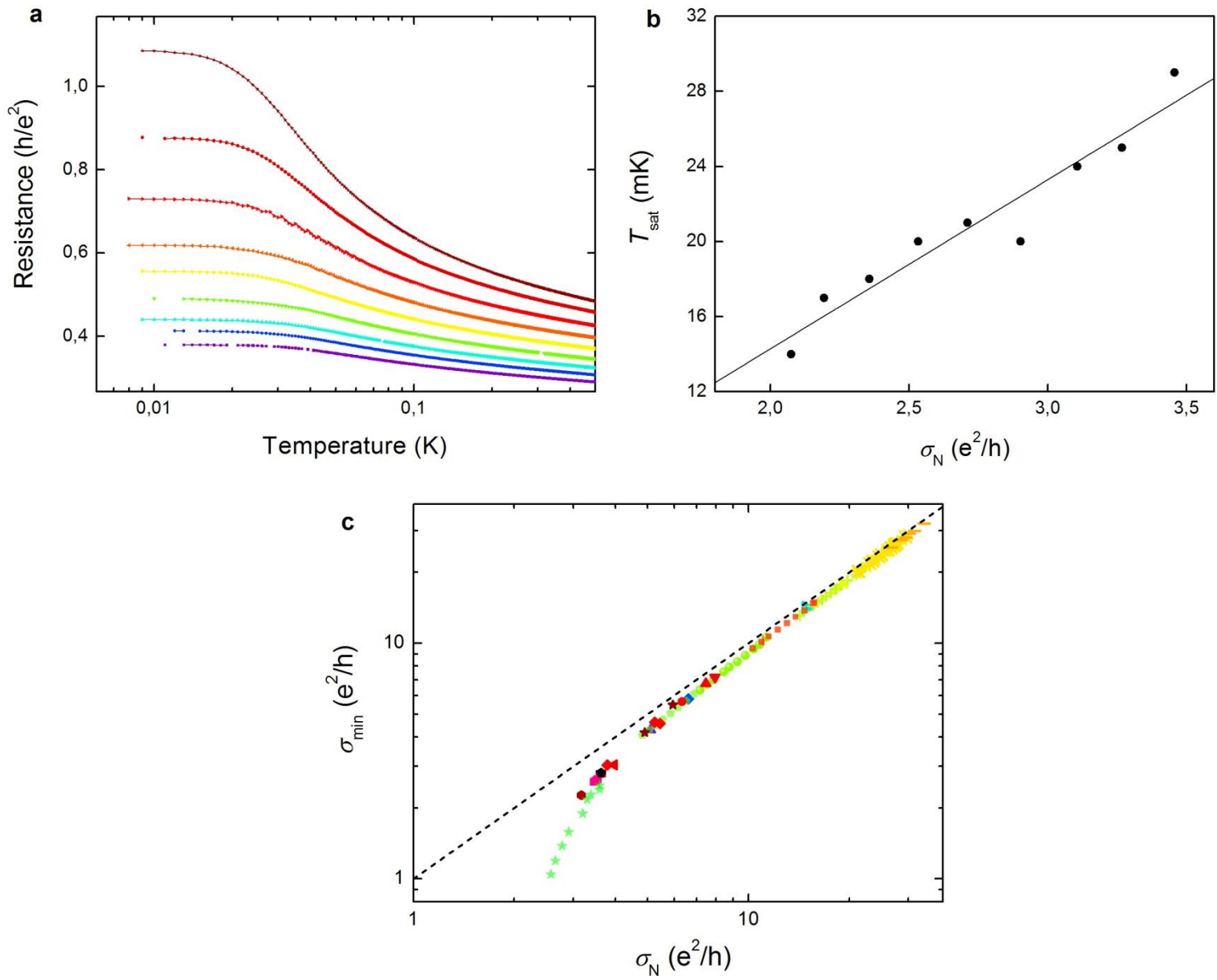

**Fig. 2**

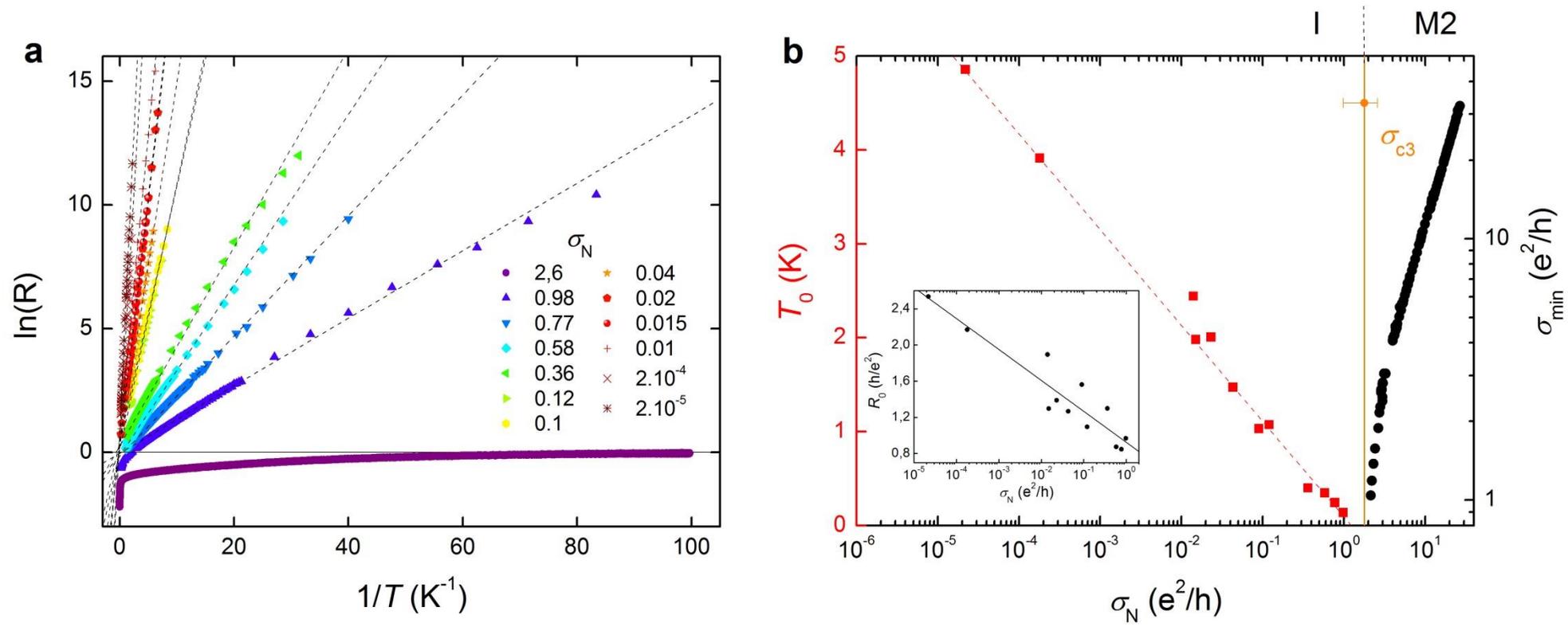

**Fig. 3**

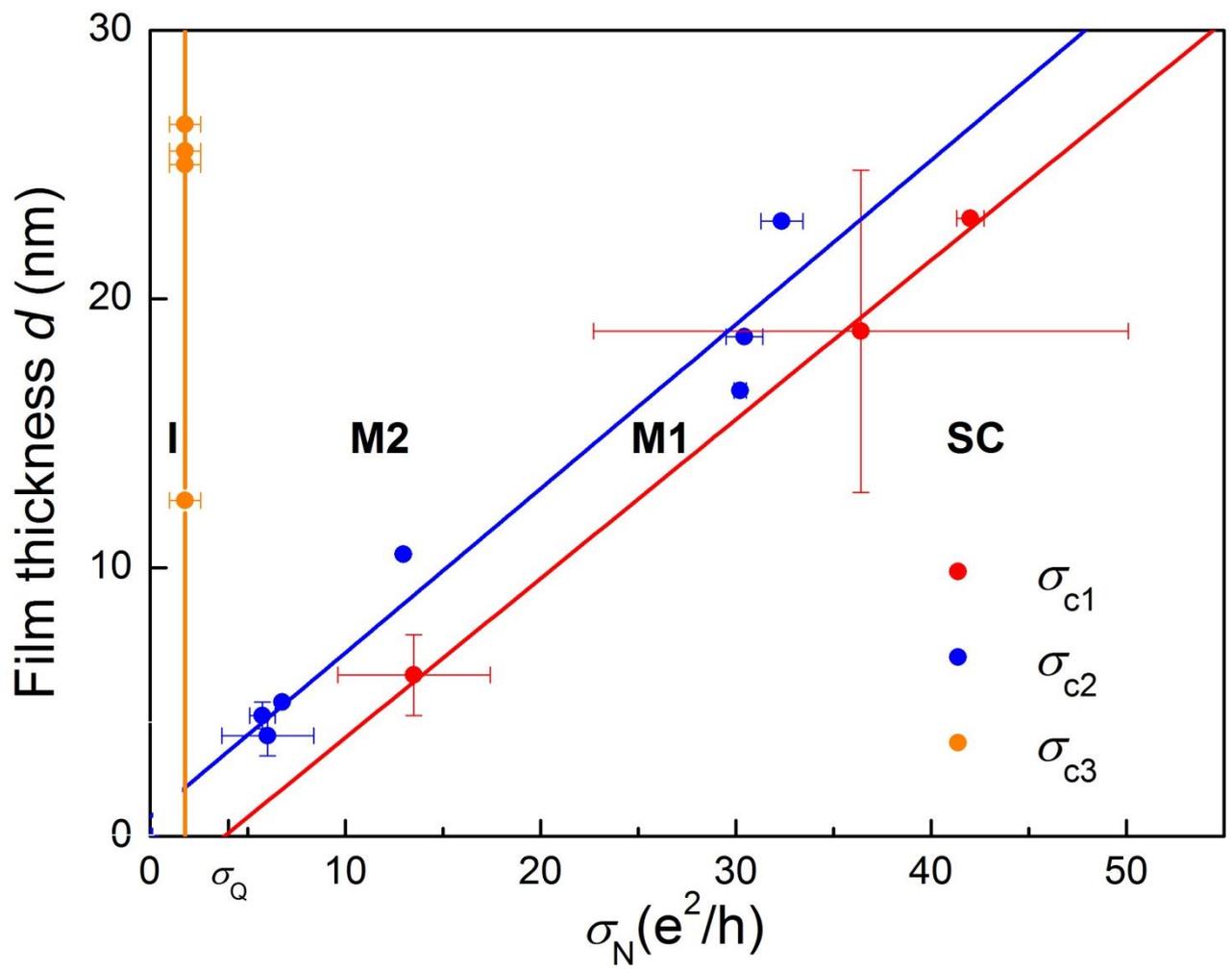

**Fig. 4**

**Supplementary materials for**

# Dissipative phases across the superconductor-to-insulator transition

F. Couëdo, O. Crauste, A.A. Drillien, V. Humbert, L. Bergé, C.A. Marrache-Kikuchi* and L. Dumoulin

CSNSM, Univ. Paris-Sud, CNRS/IN2P3, Université Paris-Saclay, 91405 Orsay, France

*Corresponding author (email: claire.marrache@csnsm.in2p3.fr)

This supplementary material includes:

- Supplementary Text
- Figure S1

Following the method described in [S1], we analyzed the resistance close to the superconducting transition using the formula:

$$R(T) = \frac{1}{1/R_{\text{ref}} + \Delta G(T)}$$

$R_{\text{ref}}$ is chosen at high temperature, typically around 10K. $\Delta G(T)$ is the quantum correction to the conductance and includes:

- the weak localization and the Coulomb interaction term $\Delta G^{\text{WL}} + \Delta G^{\text{IEE}} = G_{00} \, A \ln(\frac{T\tau k_B}{\hbar})$, where $G_{00} = \frac{e^2}{2\pi^2 \hbar}$, the coefficient A reflects the weight of the localization and $\tau$ is the elastic mean free path

- the 2D Aslamazov-Larkin term $\Delta G^{\text{AL}} = \frac{e^2}{16\hbar} \varepsilon^{-1}$ describing the formation of fluctuating Cooper pairs above $T_{c0}$, where $\varepsilon = \frac{T - T_{c0}}{T_{c0}}$

- the correction $\Delta G^{\text{DOS}} = G_{00} \ln[\frac{\ln(T_{c0}/T)}{\ln(\tau k_B T_{c0}/\hbar)}]$ due to the reduction of single electron density of states by the formation of the fluctuating Cooper pairs

- the Maki-Thompson correction $\Delta G^{MT} = \frac{e^2}{8\hbar} \frac{1}{\varepsilon - \delta} \ln(\varepsilon/\delta)$ arising from the coherent scattering of the electrons forming a Cooper pair on impurities, $\delta$ is the pair-breaking parameter

The fitting parameters here are the coefficient A, $T_{c0}$ and $\delta$.

Figure S1 shows the result of this analysis for the 23-nm-thick a-Nb$_{13.5}$Si$_{86.5}$ film, annealed at $\theta_{\text{ht}}$ = 110°C. We obtained A = 2.1, consistent with previous results on disordered thin films [S1]. $\delta$= 0.03 matches the expected value ($\delta$= 0.03), taking into account the normal state resistance $R_N$ = 646 Ω [S2]. Finally, the value of $T_{c0} = 0.058$ K agrees well with the one defined by the maximum of the derivative at low temperature ($T_{c0} = 0.06$ K).

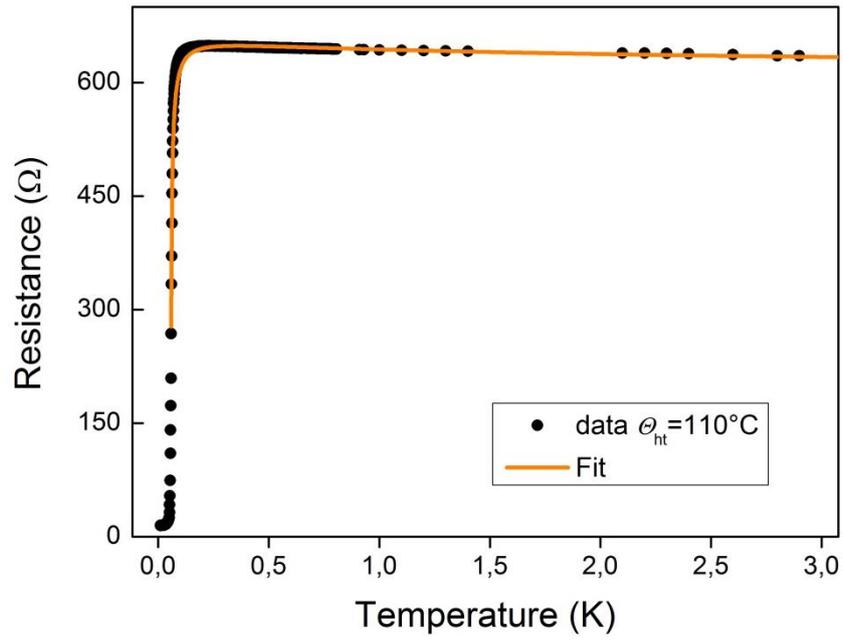

**Supplementary Figure S1.** Sheet resistance as a function of temperature for the 23-nm-thick a-Nb$_{13.5}$Si$_{86.5}$ film at the annealing temperature $\theta_{ht}$ = 110° C (Black). The orange curve shows the fitting of the quantum corrections with $A = 2.1$, $T_{c0} = 0.058$ K and $\delta = 0.03$.